# Prospects of polarized fixed target Drell-Yan experiments

**M.X. Liu[a], X. Jiang[a], D.G. Crabb[b], J.P. Chen[c], M. Bai[d]**

[a]Physics Division, Los Alamos National Laboratory, Los Alamos, NM 87545

[b]Physics Department, University of Virginia, VA 22903

[c]Thomas Jefferson National Accelerator Facility, Newport News, VA 23606

[d]Accelerator Division, Brookhaven National laborotary, Upton, NY 11973

[a]Email: mliu@lanl.gov

**Abstract**. It has been proposed that the Siverse transverse single spin asymmetry in Drell-Yan production in transversely polarized p+p collisions would have an opposite sign compared to what has been observed in the polarized Semi-Inclusive Deep Inelastic Scattering (SIDIS) experiments. Experimental confirmation or disproval of this prediction would provide a new fundamental test of QCD and shed new light on our theoretical understanding of the transverse spin physics phenomena. We discuss the prospects and physics sensitivities of polarized fixed target Drell-Yan experiments that could utilize the existing proton and other hadron beams at Fermilab, and polarized proton beams at RHIC with a polarized solid proton and/or neutron target option. We show that if realized, the new experiments would provide critical measurements of not only the sign change (or not) of Sivers functions, but also the information of quark and antiquark's Sivers distributions over a wide kinematic range.

## 1. Introduction

Large Transverse Single Spin Asymmetries (TSSAs) have been observed for decades in transversely polarized hadron reactions. Understanding the underlying physics from the first principle in QCD has been a major challenge for high-energy nuclear and particle physicists. Two fundamental mechanisms have been investigated to explain the observed large TSSAs in hard processes: 1) the Sivers mechanism [1], which generates a TSSA through the transverse distribution of quarks and gluons due to parton orbital motions, and 2) the Collins mechanism [2], which generates a TSSA through the spin dependent hadronization process of transversely polarized quarks. Measurements of significant non-zero Sivers and Collins asymmetries in polarized SIDIS at large-x region have been reported recently by the HERMES and COMPASS collaborations, as well as experiments at JLab [3,4,5]. Sivers and Collins functions are also extracted from global fitting recently [6].

In SIDIS, see the left panel in Figure 1, the final-state interaction between the struck quark (red line) and the nucleon remnant (dashed green & blue lines) is attractive due to opposite color charges. Current theoretical understanding predicts that for the Drell-Yan process in proton-proton collisions, where the incoming quark from the polarized proton annihilates with an anti-quark in the target proton, the interaction between the anti-quark (red line) and the remnant (dashed red line) of the transversely polarized proton is repulsive, see the right panel in Figure 1. As a result, the Sivers functions contribute with opposite signs to the single-spin asymmetries for these two processes [7,8,9,10].

$$\Delta^N f_{q/h\uparrow}^{SIDIS}(x, k_\perp) = -\Delta^N f_{q/h\uparrow}^{DY}(x, k_\perp)$$

This is a new fundamental prediction based on QCD factorization (where short-distance and long distance phenomena are separable in QCD calculations) and color gauge invariance. An experimental test of this prediction has become one of the top priorities for the worldwide high-energy hadronic physics community for the coming decade. All of the existing and future major high-energy nuclear physics facilities, including RHIC, the JLab 12GeV upgrade and the future EIC in US, GSI in Germany, J-PARC in Japan, have listed this physics as a major thrust of the future hadron physics programs. It tests all concepts for analyzing hard-scattering reactions that we know of today and its verification (or disproval) will be a milestone for the field of hadronic physics. Until proven, serious doubts will linger over the field of QCD-spin physics on issues such as the applicability of QCD-factorization and our understanding of the mechanisms that generates the large transverse single-spin asymmetries.

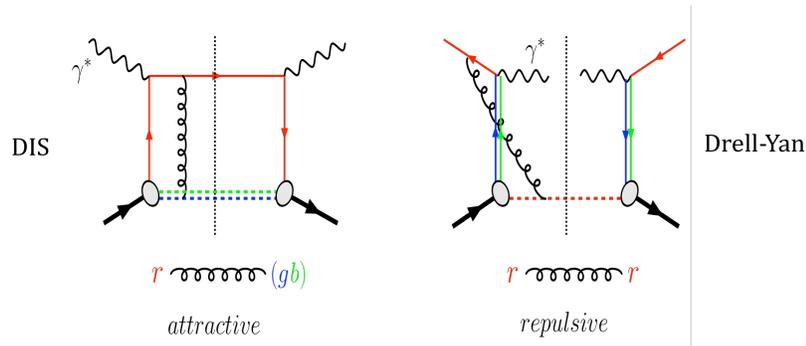

**Figure 1.** A QCD example for process-dependence of attractive interaction in DIS (left) and repulsive force in the Drell-Yan process (right). In DIS, the out-going quark (red) interacts attractively with the target remnant; in the Drell-Yan process, the in-coming antiquark (red) interacts repulsively with the polarized beam remnant.

## 2. Opportunities for polarized Drell-Yan experiments

The Drell-Yan process, $q + \bar{q} \rightarrow \gamma^* \rightarrow l^+ l^-$, has been used as a clean probe to access quark and antiquark information inside the nucleon. Theoretical understanding of Drell-Yan production in p+p collisions is believed to be robust. It is the annihilation of a quark from one proton and an antiquark from another proton to create a virtual photon in the collision of two protons at high energy, see Figure 2. Drell-Yan events are normally measured in the di-lepton (di-muon or di-electron) invariant mass range of 4<M<9 GeV where other known physics process have minimal contributions.

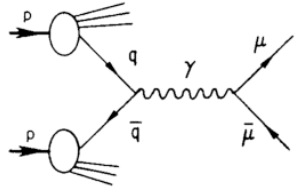

**Figure 2.** Dimuon production via Drell-Yan process in high-energy p+p collisions.

At the leading order, the Drell-Yan production cross section can be written in terms of products of beam and target parton density distributions of quark and antiquark of flavor-$i$ ($u, d, s…$):

$$\frac{d^2\sigma}{dx_1 dx_2} = \frac{4\pi\alpha^2}{9sx_1 x_2} \sum_i e_i^2 [q_i(x_1)\bar{q}_i(x_2) + \bar{q}_i(x_1)q_i(x_2)]$$

where $s$ is the square of the center of mass energy of two colliding hadorns, $e_i$ is the charge of quark-$i$, $x_1$ and $x_2$ represent the fraction of nucleon momentum carried by each of the colliding (anti)quarks and $q_i(x)$ is the corresponding quark probability density distribution. One defines the Feynman variable $x_F = x_1 - x_2$. In the forward kinematic region of $x_F >> 0$ ($x_1 >> x_2$) the cross section is dominated by contributions from beam valence quark and target antiquark while in the backward region $x_F << 0$ ($x_1 << x_2$), the cross section is dominated by beam antiquark and target valence quark contributions.

2.1. Polarized Drell-Yan experiments with polarized targets at Fermilab

The primary goal of the proposed experiment is to study the Sivers transverse single spin asymmetry and test its sign change in a broad kinematic range in Drell-Yan production using the 120 GeV unpolarized proton beams from the Main Injector at Fermilab. We study the feasibility of a polarized fixed target Drell-Yan measurement based on the Fermilab E906 experimental setup. The E906 experiment is designed to detect high mass dimuons and the primary purpose is to measure the sea quark asymmetry $\bar{d}/\bar{u}$ inside the proton in the large-x region by colliding 120 GeV on hydrogen and deuterium targets as well as to determine the fast quark energy loss in nuclear medium through p+A collisions. E906 is scheduled to take data from early 2011 for two years. We note that 120 GeV beam energy at Fermilab is low enough to allow us to access both the valence and sea quarks in the large-x region where significant spin and sea quark asymmetry effects are observed.

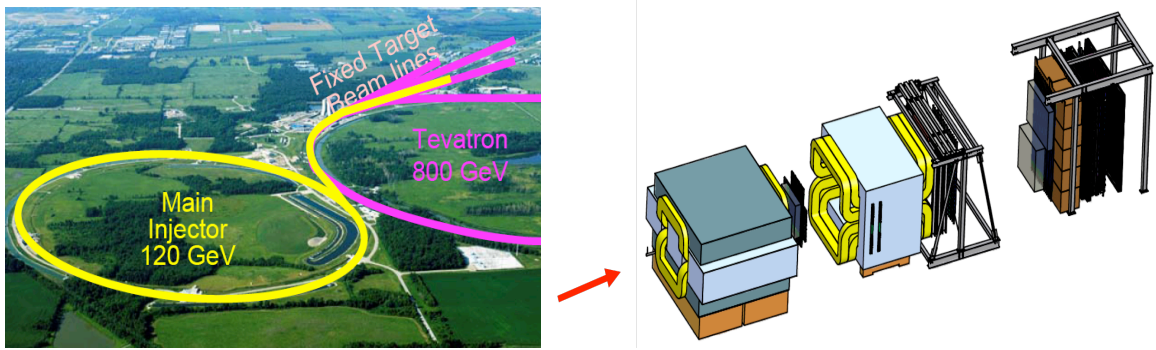

**Figure 3.** E906 dimuon experiment at Fermilab using 120 GeV proton beams from the Main Injector.

Since the beam is not polarized at Fermilab (yet), we need a polarized target at least to do spin asymmetry study. We propose to utilize the Fermilab high intensity 120 GeV main injector proton beam with a transversely polarized solid proton target ($NH_3$) to measure the left-right dimuon Drell-Yan production Sivers single transverse spin asymmetry. Significant Siver TSSAs have been observed in the target valence region in polarized SIDIS experiments. So the preferred kinematic region of target valance quarks in polarized fixed target Drell-Yan experiment corresponds to the backward produced Drell-Yan muon pairs in the center of mass system. In the laboratory frame, such event is an oppositely charged muon pair with large invariant mass but low longitudinal momentum. Fig. 3 (left panel) shows the E906 detector schematic layout. The 120 GeV proton beam comes from the left-side and hits the target, producing a dimuon through the Drell-Yan process. Only the high-energy muons are able to penetrate the thick hadron absorbers and reach the tracking detectors, while most of the background hadrons are suppressed. To maximally capture the backward produced Drell-Yan dimuons, we need to allow low longitudinal momentum muons from Drell-Yan to reach the detectors. This can be achieved by increasing the opening angle of the apparatus by placing tracking detectors closer to the target and possibly also reducing the absorber thickness. Figure 3 (right panel) shows the accepted Drell-Yan dimuons from PYTHIA simulations with simple angle and momentum selections as a function of Drell-Yan longitudinal momentum fraction $x_F$. The black curve represents the E906 acceptance with default muon minimum momentum cut, which clearly misses most of the $x_F < 0$ region. As discussed above, to test the sign change, the most interesting events are those in the region of $x_F < 0$ where significant spin asymmetries are observed in polarized SIDIS. The red curve corresponds to the events with desired much lower minimum muon momentum cut of 7.5 GeV. The main backgrounds most likely come from pion and kaon decays.

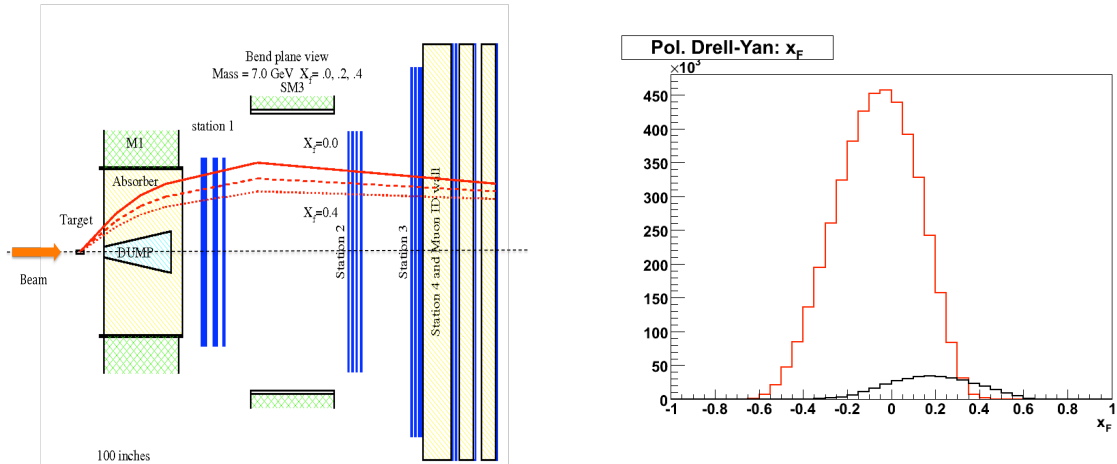

**Figure 4.** Left: E906 bend plane view of the trajectories of one of the two muons from Drell-Yan production. Right: Simple solid angle model estimate of accepted high mass Drell-Yan events (4<Mass<9 GeV) as a function of $x_F$. Red curve: proposed 4-year run with minimum muon momentum cut of $P_z > 7.5$ GeV; Black cure: 1-year E906 run with 50 cm long liquid hydrogen target.

Drell-Yan statistics is also one of the major experimental challenges. Since the $p + p \rightarrow \gamma^* \rightarrow \mu^+ \mu^-$ Drell-Yan reaction at 120 GeV has a very small cross section, a thick, high-density polarized target is critical to reach the physics goals of the proposed experiment. Initial study showed that the JLab style dynamically polarized solid ammonia ($NH_3$) target, with 6 cm in target thickness, or twice the thickness of the existing JLab target, could be realized and could be ideal for the polarized fixed target Drell-Yan experiment. The luminosity and beam energy deposition on polarized target at Fermilab Drell-Yan experiment will be similar to the ones at JLab Hall-C high intensity electron scattering experiments. To maintain the target polarization, the method of dynamic polarization relies on the strong magnetic field acting on magnetic moments at a low temperature to generate a population difference that favors one spin state. The existing JLab Hall-C polarized target was built two decades ago and was used in many experiments at SLAC and JLab. In this target, closely packed solid ammonia beads are immersed in liquid $^4$He and maintained at 1°K with a $^4$He evaporation refrigerator. The target polarization was routinely kept at 80% at JLab experiments and the target spin direction was reversed once every eight hours. Target annealing (warm-up) was required once every few days in order to maintain a high polarization.

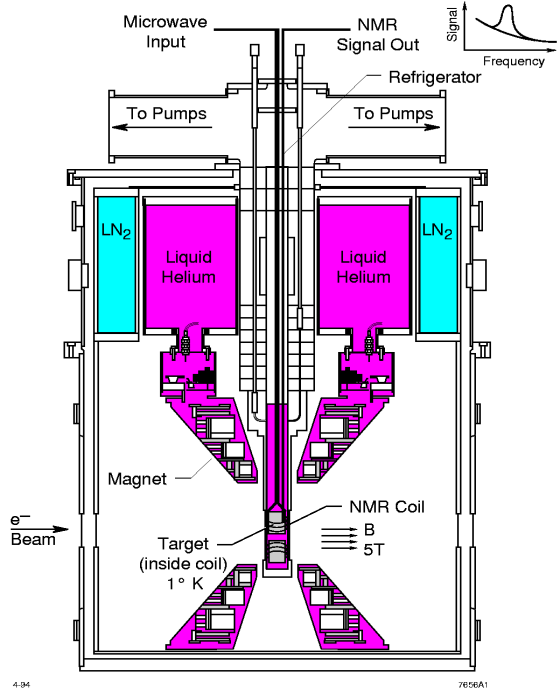

**Figure 5.** Diagram of the UVA/SLAC/JLab polarized target being used at JLab Hall-C. The superconducting magnet produces a highly uniform 5 Tesla field over the target volume. At a temperature of 1K, $^4$He evaporation refrigerator provides about 1 W of cooling power to the irradiated ~2mm diameter NH$_3$ beads inside the target cell.

Our preliminary experimental sensitivity studies are based on the assumptions of a new 6 cm long polarized solid NH$_3$ target, with a polarization of 80% and a dilution factor (polarized nucleon vs. all nucleon) of 0.22. We also assume four six-month runs with a proton flux of $10^{13}$ protons/minute (for a total of $10^{19}$ protons) at 120 GeV. It is worth to note that besides testing the sign change in the Sivers asymmetry for the valence quarks, the proposed experiment would also provide precision measurements of sea-quark's Sivers function at large-x that is impossible to do in polarized SIDIS experiments. According to some models, such as meson cloud models, sea quark (antiquark) could have significant none-zero Sivers functions at large x. So it would be very interesting to measure antiquarks' Sivers functions at large-x, corresponding to the region of $x_F > 0$ on the right-hand plot in Figure 6.

We note that there is an ongoing effort to polarized the 120GeV proton beam from the main injector at Fermilab [11]. If realized, it will allow us to expand the physics program significantly by including double spin asymmetry measurements.

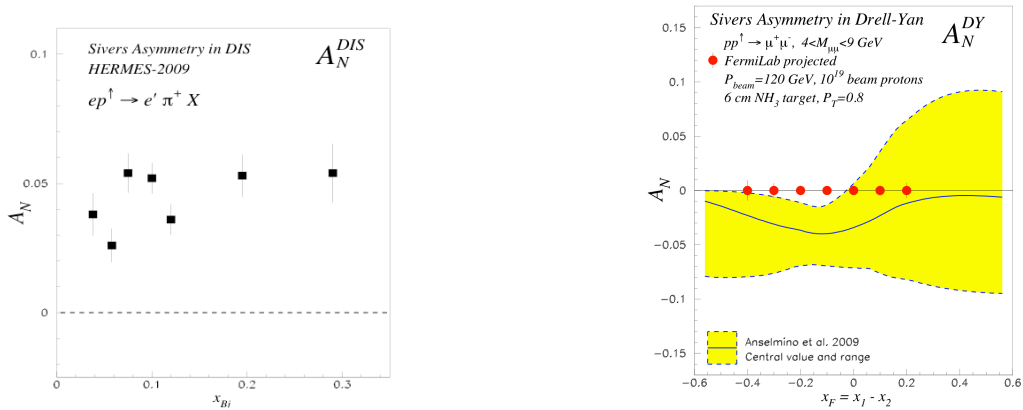

**Figure 6.** Left: Sivers asymmetry in DIS measured by the HERMES for $\pi^+$ production. Right: predictions of Drell-Yan single spin asymmetry and the expected statistical uncertainties of the proposed experiment at Fermilab. Note that poorly known sea-quark Sivers functions could also be determined precisely at large-x region (corresponding to the $x_F>0$ region).

2.2. Polarized Drell-Yan fixed target experiments with polarized beams at RHIC

The Relativistic Heavy Ion Collider at Brookhaven National Laboratory is the world only high-energy polarized proton collider. A feasibility study of Drell-Yan TSSA measurements in the collider mode was carried out in 2007 using PHENIX and STAR detectors and the results are very favorable [13]. One of the major experimental challenges is the required high luminosity since the Drell-Yan production cross section is very small. Significant improvement in luminosity is needed for the future runs to achieve the goal. At present, the yield is too low for any meaningful Drell-Yan asymmetry measurement.

Here we explore a fixed target Drell-Yan option with the polarized 250 GeV proton beam at RHIC, where the target could be internal or external ones. Compared to the collider mode where only about 1% of protons in a store used for collisions, a fixed target experiment could use most of the polarized protons from RHIC in the beam dump mode. For the external target option, the target itself could also be polarized, thus allowing double spin asymmetry measurements. However this option requires adding a new slow extraction beam line to the RHIC ring in order to send polarized beams to the fixed target area. For internal target, it could be a high-density cluster-jet or pellet, or a solid (polarized frozen) target placed at certain distance from the center of the beam. Several options are being studied right now.

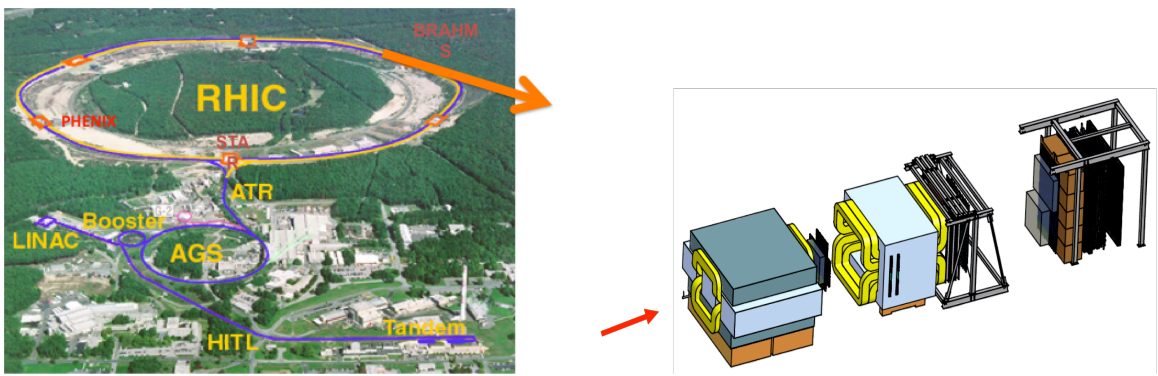

**Figure 7.** Proposed external fixed target Drell-Yan experiment with extracted polarized beams from RHIC. The same Fermilab E906 dimuon spectrometer is used for the feasibility study. Polarized target could be used for high luminosity double spin physics program.

For a dedicated fixed target experiment with E906-type setup for dimuons, we could achieve significant luminosity of the order of several 10 fb$^{-1}$ within a few years' runs. The left-hand plots in Figure 7 show the kinematic coverage of $x_{beam}$ v.s. $x_{target}$ of Drell-Yan events with different minimum muon momentum cuts, the right-hand plot shows the expected experimental sensitivity of the Drell-Yan Sivers asymmetry. The dashed lines are the theoretical bounds of the asymmetry based on recent global fits of the Sivers functions from the polarized SIDIS data.

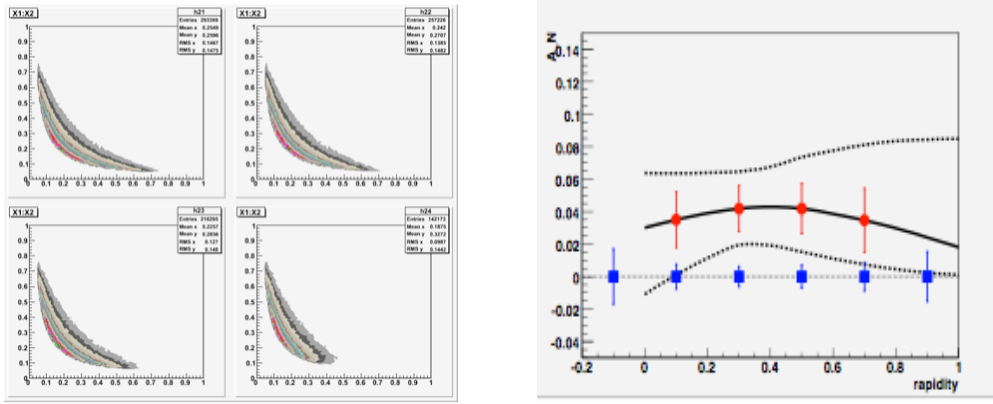

**Figure 7.** Left: the $x_{beam}$ vs $x_{target}$ coverage of Drell-Yan events with different minimum momentum cuts: top-left (p>0 GeV); top-right (p> 5GeV); bottom-left (p>10GeV); Right: Statistical sensitivity to the Drell-Yan Siver asymmetry with two expected luminosities: L=10 fb$^{-1}$(red); L=50 fb$^{-1}$(blue). The theoretical prediction is based on the latest global fit of Sivers functions [6].

2.3. Other spin physics programs

If both the beam and target could be polarized vertically and also horizontally, it would open up a whole new opportunity for spin physics. For instance, the poorly known quark transversity distribution could be measured precisely via the double transverse spin asymmetry $A_{TT}$. Double longitudinal asymmetry $A_{LL}$ could be used to study antiquark polarization inside the polarized proton. Many of the proposed topics for Drell-Yan process could also benefit from J/Psi and open charm measurements. Unlike the Drell-Yan, which is an electromagnetic process, J/Psi and open charm productions are strong interaction processes involving gluon-gluon and quark-antiquark fusions. Comparison between Drell-Yan and J/Psi (or open charm) production will further elucidate various aspects of parton distributions and QCD dynamics.

**3. Summary**

Our preliminary study shows that a polarized fixed target Drell-Yan experiment at Fermilab and/or RHIC with polarized target and/or beam would provide a unique program for spin physics complementary to polarized DIS experiments and the current RHIC-Spin program. Specific physics topics include the measurements of T-odd Sivers (and Boer-Mulder) distribution functions and test the new QCD prediction of the sign change, and the transversity and helicity distributions of partons if both beam and target are polarized. Particularly the expected high luminosity and the broad kinematic coverage of x range in fixed target experiments would allow us to carry out unique precision measurements that is hard to achieve in other approaches. Such experiment is also ideal for the study of sea quark and gluon distributions at large-x.

**References**
[1]   D.W. Sivers, Phys. Rev. D **41**, 83 (1990).
[2]   J.C. Collins, Nucl. Phy. Lett. B **536**, 43 (2002).

[3]     HERMES collaboration, Phys. Rev. Lett. **94**, 012002(2005); Phys. Rev. Lett. **103** (2009) 152002.
[4]     COMPASS collaboration, Phys. Rev. Lett. **94**, 202002 (2005); Phys. Lett. B **692** (2010) 240-246.
[5]     JLab neutron transversity experiment E06-10.
[6]     M. Anselmino *et al*., Eur. Phys. J. A **39**, 89 (2009).
[7]     S. J. Brodsky, D. S. Hwang and I. Schmidt, Phys. Lett. B **530**, 99 (2002); Nucl. Phys. B **642**, 344 (2002).
[8]     J. C. Collins, Phys. Lett. B **536**, 43 (2002).
[9]     X. Ji and F. Yuan, Phys. Lett. B **543**, 66 (2002); A. V. Belitsky, X. Ji and F. Yuan, Nucl. Phys. B **656**, 165 (2003).
[10]    D. Boer, P. J. Mulders and F. Pijlman, Nucl. Phys. B **667**, 201 (2003).
[11]    A. Krisch, W. Lorenzon *et al*., Santa Fe Polarized Drell-Yan Workshop, 10/31-11/1, 2010, http://p25ext.lanl.gov/~ming/SantaFe-DY/Drell-Yan-Workshop.htm
[12]    RHIC-SPIN 2007 whitepaper: Transverse Single Spin Asymmetry in Drell-Yan Production in proton+proton collisions, http://spin.riken.bnl.gov/rsc/write-up/dy_final.pdf